\documentclass[prd,preprint,preprintnumbers,nofootinbib,eqsecnum,superscriptaddress]{revtex4}

 \usepackage[dvips,final]{graphicx}
  \usepackage{amssymb}
   \usepackage{amsmath}
    \usepackage{amsfonts}
     \usepackage{epsfig}
      \usepackage{bm}

\usepackage{mathpazo}

\usepackage[section]{placeins}

\input epsf.tex
\def\desepsf(#1 width #2){\epsfxsize=#2 \epsfbox{#1}}

\newcommand{\alphas}{\ensuremath{\alpha_\mathrm{s}}}

\newcommand{\PBM}{PB}

\usepackage{multirow}
\usepackage{ctable}
\usepackage{booktabs}
\usepackage{array}
\usepackage{tabularx}
\usepackage{xcolor}
\usepackage{pstricks}
\definecolor{fred}{rgb}{0.90053, 0.00369, 0.00159}  

\newcommand{\dif}{\mathrm{d}}
\newcommand{\diff}[1]{\frac{\mathrm{d}#1}{#1}}

\newcommand{\be}{\begin{eqnarray}}
\newcommand{\ee}{\end{eqnarray}}

\begin{document}

\author{Rafa{\l} Maciu{\l}a}
\email{rafal.maciula@ifj.edu.pl}
\affiliation{Institute of Nuclear
Physics, Polish Academy of Sciences, ul. Radzikowskiego 152, PL-31-342 Krak{\'o}w, Poland}



\title{QCD predictions for open charm meson production \\at the LHCb in fixed-target experiment}

\begin{abstract}
We investigate open charm meson production in fixed-target LHCb experiment at $\sqrt{s}=86.6$ GeV in $p+^4\!\mathrm{He}$ collisions.
Theoretical calculations of charm cross section are done in the framework of the $k_{T}$-factorization approach.
Its application in the kinematical range never examined before is carefully discussed. 
We consider different schemes for the calculations relevant for different unintegrated (transverse momentum dependent) parton densities in a proton.
We include in the analysis both CCFM- and DGLAP-based models of unintegrated parton distributions appropriate for the considered kinematics. 
Integrated as well as differential cross sections as a function of $D^{0}$ meson rapidity and transverse momentum are shown
and compared with the experimental data. As a reference point, predictions of next-to-leading order collinear approach are also presented and discussed. A very good agreement between the experimental data and the $k_{T}$-factorization predictions was obtained.
Both the CCFM and the DGLAP-based frameworks for parton distributions in a proton are successfully used to explain the LHCb fixed-target open charm cross section.   
\end{abstract} 

\maketitle

\section{Introduction}

According to current knowledge and experimental abilities, production of heavy flavour is known as the best testing ground to investigate foundations of the theory of hard QCD interactions. Theoretical studies of \textit{e.g.} charm cross section in proton-proton collisions provide an unique precision tool in this context. Phenomenology of charm particles production at hadron colliders has been shown many times to be one of the most powerful tools in testing pQCD techniques.

In the ongoing LHC era the activity in this context has increased significantly on both experimental and theoretical sides, since charm particles
are copiously produced at currently available high energies.
Various measurements of open charm meson production have been accomplished by the ALICE, ATLAS, CMS and LHCb experiments in $pp$, $pA$ and $AA$ reactions at different $\mathrm{TeV}$-scale energies (from 2.76 to 13 TeV). During last years several phenomenological studies of open charm production at the LHC were performed, including inclusive $D$-meson (see \textit{e.g.} Refs.\cite{Cacciari:2012ny,Maciula:2013wg,Klasen:2014dba}) and $\Lambda_c$-baryon \cite{Maciula:2018iuh} production, $D \overline D$ meson-antimeson pair production \cite{Maciula:2013wg,Karpishkov:2016hnx}, double \cite{Maciula:2013kd} and even triple \cite{Maciula:2017meb} charm production, as well as associated charm production with jets \cite{Maciula:2016kkx,Maciula:2017egq} and gauge bosons \cite{Lipatov:2019izq}.  

Very recently the LHCb collaboration has performed a first measurement of the charm meson cross section in fixed-target configuration \cite{Aaij:2018ogq}. Both, production of hidden ($J\!/\!\psi$) and open ($D^{0}$) charm was carefully studied in fixed-target $p+^4\!\mathrm{He}$ and $p+^{40}\!\!\mathrm{Ar}$ collisions
at $\sqrt{s_{NN}}=86.6$ and $\sqrt{s_{NN}}=110.4$ GeV, respectively. The absolute cross sections (integrated and differential) were reported in the case of the $p+^4\!\mathrm{He}$ interactions only.

Here we wish to make theoretical analysis of the LHCb fixed-target charm data and to examine the $k_{T}$-factorization approach \cite{kTfactorization} in this context, so far not explored in this unique kinematical regime.
This approach allows very easily to include higher-order QCD radiative corrections (namely, part of NLO + NNLO + ... terms corresponding to real initial-state gluon emissions) that can be taken into account via the so-called unintegrated (transverse momentum dependent) parton distribution functions (uPDFs) in a proton.

The $k_{T}$-factorization has become a widely exploited tool and it is of common interest and importance to test it in many different processes and in various kinematical regimes. The kinematical configuration of the fixed-target LHCb experiment corresponds to the region where both the Catani-Ciafaloni-Fiorani-Marchesini (CCFM) \cite{CCFM} and the Dokshitzer-Gribov-Lipatov-Altarelli-Parisi (DGLAP) \cite{DGLAP} evolution equations are legitimated for any pQCD theoretical calculations and could in principle be used to describe the dynamics behind the mechanisms of \textit{e.g.} open charm meson production. The LHCb fixed-target open charm data brings therefore a new opportunity to test the CCFM- and the DGLAP-based unintegrated (transverse momentum dependent) parton distributions (uPDFs) in a proton. Here, the uPDFs can be probed at rather intermediate (and even large) longitudinal momentum fractions $x$ where in turn effects related to the Balitsky-Fadin-Kuraev-Lipatov (BFKL) \cite{BFKL} evolution should not appear.

The BFKL equation resums large logarithmic terms proportional to $(\alpha_{s} \ln s)^{n} \sim (\alpha_{s} \ln 1/x)^{n}$, important at high energies and small-$x$. The CCFM equation takes into account additional terms proportional to $(\alpha_{s} \ln 1/(1-x))^{n}$ and is valid at both low and large $x$. 
Some methods to calculate the uPDFs from the conventional DGLAP equations are also present in the literature. An application of the different models of uPDFs in phenomenological studies of a given process is not automated as in the case of the collinear PDFs. Here a deliberate adjustment of the model calculations to theoretical concepts contained in the construction of a chosen uPDF is required.    
 
Very, recently a new scheme for a calculation of partonic cross section in the $k_{T}$-factorization
approach has been proposed in Ref.~\cite{Maciula:2019izq}. There, a new calculation scenario where higher-order QCD hard radiative corrections are not resummed in the uPDF but are taken into account via tree-level hard matrix elements was used to calculate charm production at $7$ TeV in $pp$-collisions in the LHCb experiment. Then, the same ideas were also successfully used for the associated production of electroweak gauge bosons and heavy quark jets at the LHC \cite{Lipatov:2019izq}.
       
Here we wish to investigate the correspondence between different scenarios of charm cross section calculation in the $k_{T}$-factorization approach relevant for the CCFM uPDFs \cite{Hautmann:2013tba,Jung:2004gs} as well as for the (DGLAP-based) Parton-Branching (PB) \cite{Martinez:2018jxt} and Kimber-Martin-Ryskin (KMR) \cite{Kimber:1999xc,Kimber:2001sc,Watt:2003mx} uPDFs. One of the main goals of this paper is to compare predictions of the different scenarios and to find a similarity and connection between them.

\section{Details of the model calculations}

\subsection{Cross section for charm quark and meson production}

In this subsection we wish to shortly describe two different schemes of the calculations of the $c\bar c$-pair production cross section
within the $k_{T}$-factorization approach:
\begin{itemize}
\item the standard model with the leading-order $2\to 2$ matrix elements and with extra hard emissions from the uPDF (see \textit{e.g.} Ref.~\cite{Maciula:2013wg} and references therein),
\item the new model with the higher-order $2 \to 3$ and $2 \to 4$ matrix elements and without resummation of extra hard emissions in the uPDF (see Ref.~\cite{Maciula:2019izq}).
\end{itemize} 

The standard model of the calculation can be successfully used in phenomenological studies only with the uPDFs that effectively take into account higher-order contributions. Technically, it means that the transverse momentum $k_{t}^{2}$ generated in the evolution of the uPDF is allowed to be larger than the scale $\mu_{F}^2$, which is in principle the case of the CCFM unintegrated distributions. 

On the other hand, when the extra hard emissions from the uPDF are suppressed due to its theoretical construction, the higher-order radiative corrections has to be taken into account at the level of hard-matrix elements. This is especially the case of the DGLAP-based unintegrated parton distributions which for conclusive phenomenological studies needs to be applied within the new scheme of the $k_{T}$-factorization calculations.

\subsubsection{The standard $k_{T}$-factorization calculations within the leading-order $2\to 2$ mechanism}

\begin{figure}[!h]
\centering
\begin{minipage}{0.3\textwidth}
  \centerline{\includegraphics[width=1.0\textwidth]{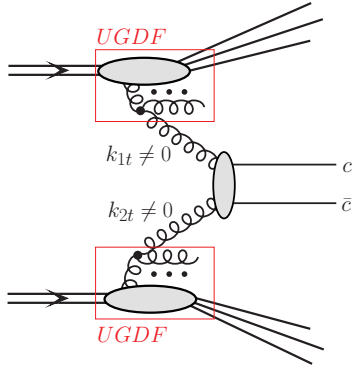}}
\end{minipage}
  \caption{
\small A diagramatic representation of the leading-order mechanism of charm production.
}
\label{fig:diagramLO}
\end{figure}

We remind the standard theoretical formalism for the calculation of the $c\bar{c}$-pair production in the $k_{T}$-factorization approach. In this framework the transverse momenta $k_{t}$'s (virtualities) of both partons entering the hard process are taken into account, both in the matrix elements and in the parton distribution functions. Emission of the initial state partons is encoded in the transverse-momentum-dependent (unintegrated) uPDFs. In the case of charm flavour production the parton-level cross section is usually calculated via the $2\to 2$ leading-order $g^*g^* \to c\bar c$ fusion mechanism of off-shell initial state gluons that is the dominant process at high energies. Even at lower energies as long as small transverse momenta and central rapidities are considered the $q^*\bar q^* \to c\bar c $ mechanism remains subleading. Then the hadron-level differential cross section for the $c \bar c$-pair production, formally at leading-order, reads:
\begin{eqnarray}\label{LO_kt-factorization} 
\frac{d \sigma(p p \to c \bar c \, X)}{d y_1 d y_2 d^2p_{1,t} d^2p_{2,t}} &=&
\int \frac{d^2 k_{1,t}}{\pi} \frac{d^2 k_{2,t}}{\pi}
\frac{1}{16 \pi^2 (x_1 x_2 s)^2} \; \overline{ | {\cal M}^{\mathrm{off-shell}}_{g^* g^* \to c \bar c} |^2}
 \\  
&& \times  \; \delta^{2} \left( \vec{k}_{1,t} + \vec{k}_{2,t} 
                 - \vec{p}_{1,t} - \vec{p}_{2,t} \right) \;
{\cal F}_g(x_1,k_{1,t}^2,\mu_{F}^2) \; {\cal F}_g(x_2,k_{2,t}^2,\mu_{F}^2) \; \nonumber ,   
\end{eqnarray}
where ${\cal F}_g(x_1,k_{1,t}^2,\mu_{F}^2)$ and ${\cal F}_g(x_2,k_{2,t}^2,\mu_{F}^2)$
are the gluon uPDFs for both colliding hadrons and ${\cal M}^{\mathrm{off-shell}}_{g^* g^* \to c \bar c}$ is the off-shell matrix element for the hard subprocess.
The gluon uPDF depends on gluon longitudinal momentum fraction $x$, transverse momentum
squared $k_t^2$ of the gluons entering the hard process, and in general also on a (factorization) scale of the hard process $\mu_{F}^2$.
The extra integration is over transverse momenta of the initial
partons. Here, one keeps exact kinematics from the very beginning and additional hard dynamics coming from transverse momenta of incident partons. Explicit treatment of the transverse momenta makes the approach very efficient in studies of correlation observables. The two-dimensional Dirac delta function assures momentum conservation. The gluon uPDFs must be evaluated at longitudinal momentum fractions 
$x_1 = \frac{m_{1,t}}{\sqrt{s}}\exp( y_1) + \frac{m_{2,t}}{\sqrt{s}}\exp( y_2)$, and $x_2 = \frac{m_{1,t}}{\sqrt{s}}\exp(-y_1) + \frac{m_{2,t}}{\sqrt{s}}\exp(-y_2)$, where $m_{i,t} = \sqrt{p_{i,t}^2 + m_c^2}$ is the quark/antiquark transverse mass.  

The off-shell matrix elements are known explicitly only in the LO and only for limited types of QCD $2 \to 2$ processes (see \textit{e.g.} heavy quark \cite{Catani:1990eg}, dijet \cite{Nefedov:2013ywa}, Drell-Yan \cite{Nefedov:2012cq} mechanisms). 
Some first steps to calculate NLO corrections in the $k_{T}$-factorization framework have been tried only very recently for diphoton production \cite{Nefedov:2015ara,Nefedov:2016clr}.
There are ongoing intensive works on construction of the full NLO Monte Carlo generator for off-shell initial state partons that are expected to be finished in near future \cite{private-Hameren}. Another method for calculation of higher multiplicity final states is to supplement the QCD $2 \to 2$ processes with parton shower. For the off-shell initial state partons it was done only with the help of full hadron level Monte Carlo generator CASCADE \cite{Jung:2010si}. However, in the moment this method can be consistently used only with the uPDFs that have a steep drop of the parton densities at $k_{t}^{2} > \mu_{F}^{2}$.

On the other hand, the popular statement is that actually in the $k_{T}$-factorization approach already at leading-order some part of radiative higher-order corrections can be effectively included via uPDFs. However, it is true only for those uPDF models in which extra emissions of soft and even hard partons are encoded, including $k_{t}^{2} > \mu_{F}^{2}$ configurations. Then, when calculating the charm production cross section via the $g^* g^* \to c \bar c$ mechanism one could expect to effectively include contributions related to an additional one or two (or even more) extra partonic emissions which in some sense plays a role of the initial state parton shower.

\subsubsection{A new scheme of the calculations with the higher-order $2\to 3$ and $2\to 4$ mechanisms}

Now we wish to shortly describe an alternative scheme of the calculation of the heavy flavour cross sections in the $k_{T}$-factorization approach, recently proposed and discussed in Ref.~\cite{Maciula:2019izq}. The main idea is to include usual leading order subprocesses properly matched with a number of additional higher-order radiative corrections at the level of hard matrix elements. This procedure is devoted in principle to the calculations for which the DGLAP-based unintegrated parton densities are applied. Here the extra hard emissions from the uPDFs are usually strongly suppressed what leaves a room for higher-order terms.   
   
Due to the lack of the full NLO and/or NNLO framework of the $k_{T}$-factorization, within the present methods the higher-order pQCD calculations can be done only at tree-level. Within the proposed scheme the $2\to 2$, $2\to 3$ and even $2\to 4$ contributions to heavy quark-antiquark pair production are summed up together under a special conditions introduced to avoid a possible double-counting (see a discussion of the double-counting-exclusion cuts in Ref.~\cite{Maciula:2019izq}).

\begin{figure}[!h]
\centering
\begin{minipage}{0.3\textwidth}
  \centerline{\includegraphics[width=1.0\textwidth]{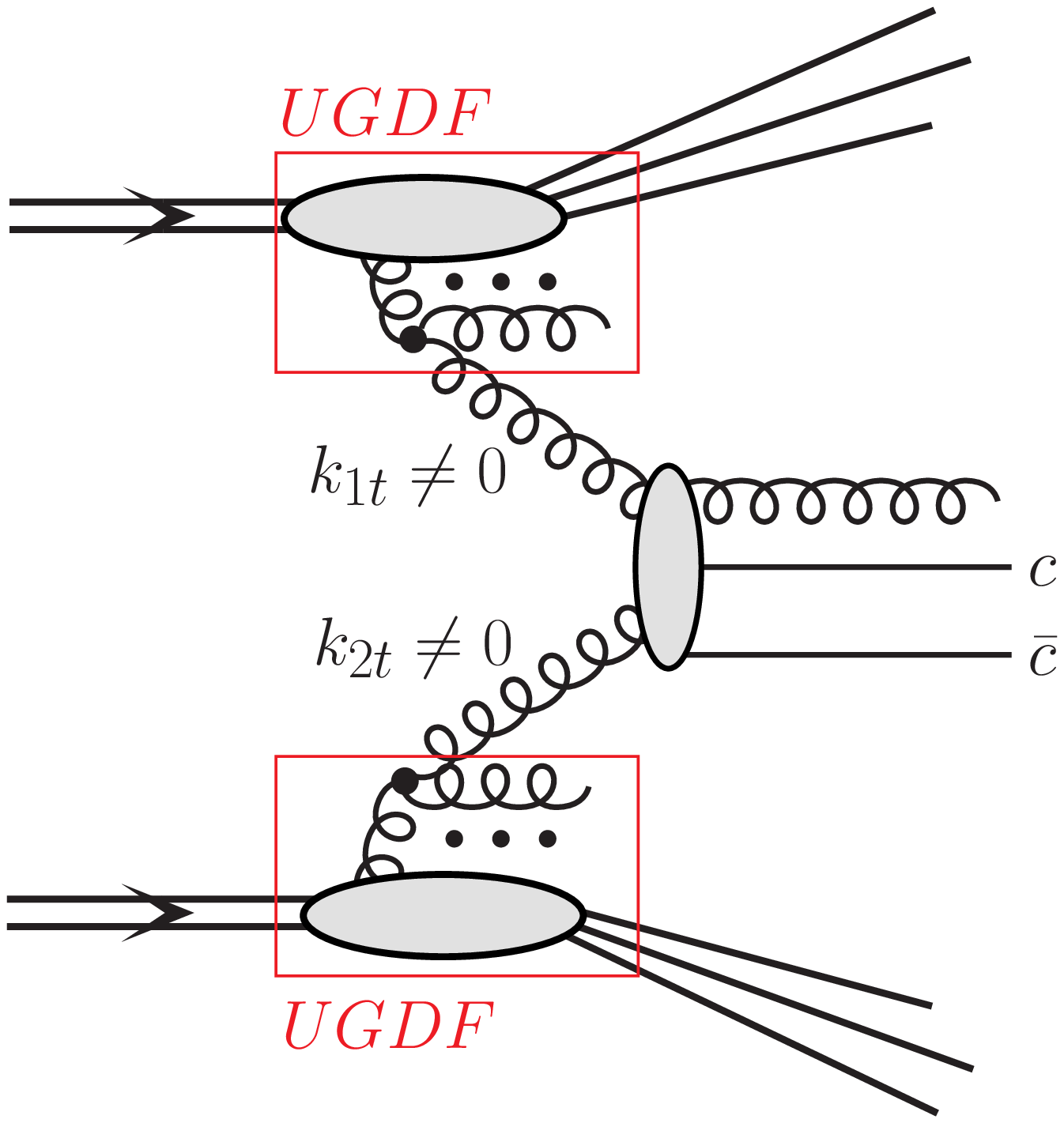}}
\end{minipage}
\begin{minipage}{0.3\textwidth}
  \centerline{\includegraphics[width=1.0\textwidth]{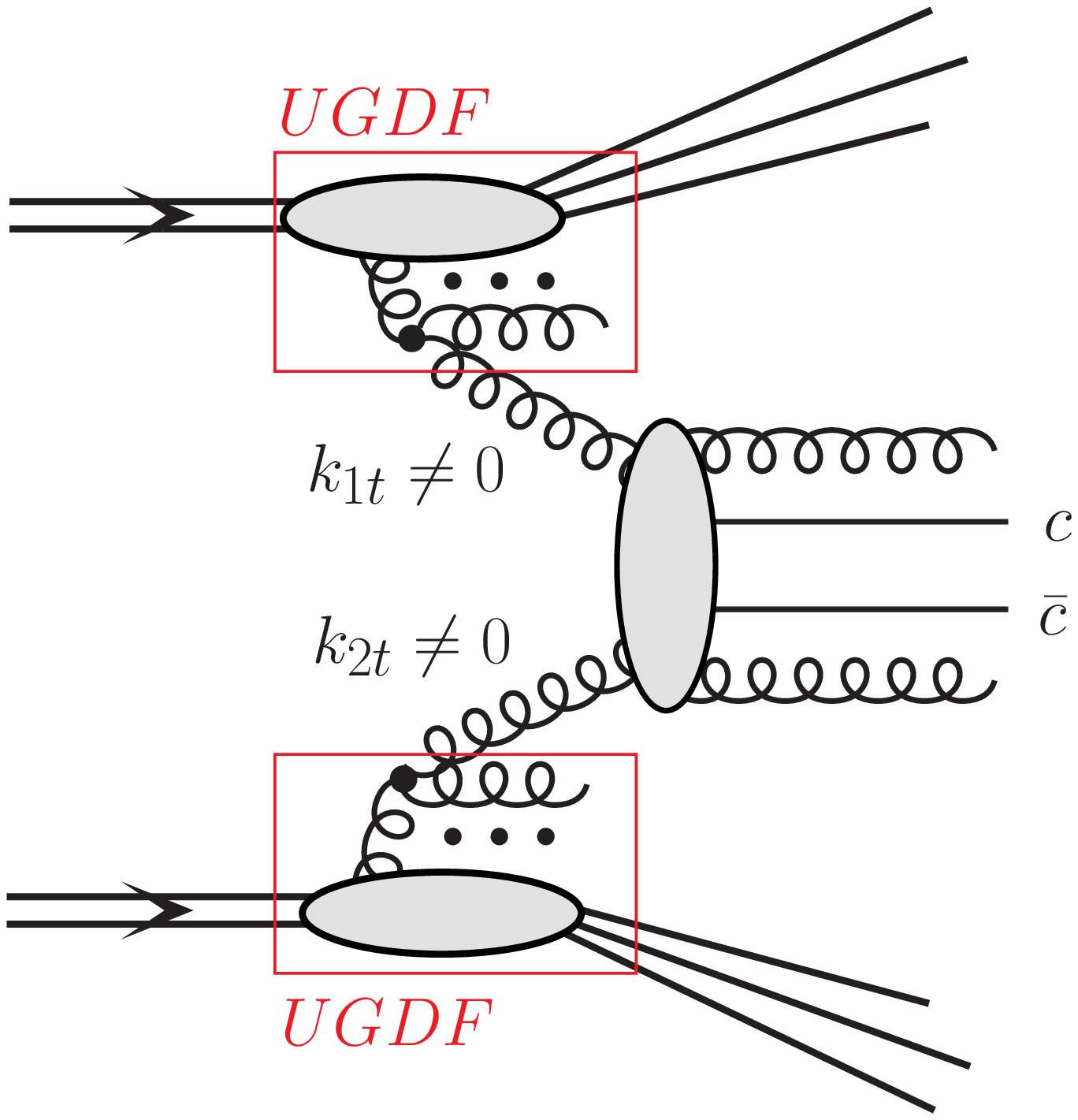}}
\end{minipage}
  \caption{
\small A diagramatic representation of an example of the higher-order mechanisms of charm production.
}
\label{fig:diagramHO}
\end{figure}

The numerical calculations for the higher-order contributions are also performed in the framework of the $k_{T}$-factorization approach within the methods adopted in the KaTie Monte Carlo generator \cite{vanHameren:2016kkz}, where the off-shell matrix elements for higher final state parton multiplicities at the tree-level are calculated numerically with the help of methods of numerical BCFW recursion 
\cite{Bury:2015dla}. We include all the possible $2\to 3$ and $2\to 4$ channels for the $c\bar c$-pair production with off-shell gluons and light quarks in the initial states. These two classes of higher-order processes are schematically illustrated in Fig.~\ref{fig:diagramHO} where Feynmann diagrams for the $g^*g^* \to g c \bar c$ and the $g^*g^* \to ggc \bar c$ mechanisms are shown as an example. 

In general, the cross secton for $pp \to g(g)c\bar c\, X$ reaction in the $k_T$-factorization approach can be written as
\begin{equation}
d \sigma_{p p \to g(g)c\bar c \; X} =
\int d x_1 \frac{d^2 k_{1t}}{\pi} d x_2 \frac{d^2 k_{2t}}{\pi}
{\cal F}_{g}(x_1,k_{1t}^2,\mu_F^2) {\cal F}_{g}(x_2,k_{2t}^2,\mu_F^2)
d {\hat \sigma}_{g^*g^* \to g(g)c\bar c}
\; .
\label{cs_formula}
\end{equation}
Then, the elementary cross section from the above can be written
somewhat formally as:
\begin{equation}
d {\hat \sigma}_{gg \to g^*g^* \to g(g)c\bar c } =
\prod_{l=1}^{n}
\frac{d^3 p_l}{(2 \pi)^3 2 E_l} 
(2 \pi)^n \delta^{n}(\sum_{l=1}^{n} p_l - k_1 - k_2) \times\frac{1}{\mathrm{flux}} \overline{|{\cal M}_{g^*g^* \to g(g)c\bar c}(k_{1},k_{2})|^2}
\; ,
\label{elementary_cs}
\end{equation}
with $n=3$ and $n=4$ for $g^*g^* \to gc\bar c$ and $g^*g^* \to ggc\bar c$, respectively, where $E_{l}$ and $p_{l}$ are energies and momenta of final state gluon(s) and charm quarks. Above only dependence of the matrix element on four-vectors of incident partons $k_1$ and $k_2$ is made explicit. In general, all four-momenta associated with partonic legs enter.
Also in this case, the matrix element takes into account that both partons entering the hard
process are off-shell with virtualities $k_1^2 = -k_{1t}^2$ and $k_2^2 = -k_{2t}^2$.

Having minijet(s) in the final state at tree-level requires some technical methods for regularization of the cross section. Here we follow exactly the method adopted in Ref.~\cite{Maciula:2019izq} which was originally proposed \textit{e.g.} in \textsc{Pythia} Monte Carlo generator \cite{Sjostrand:2014zea} for the calculations of the $2\to 2$ pQCD processes with light quarks and gluons in the final states. There, a special suppression factor $F_{\mathrm{sup}}(p_{T}) = p_{T}^4/(p^{2}_{T0}+p_{T}^2)^2$ was introduced with $p_{T}$ being the outgoing minijet transverse momentum and $p_{T0}$ being a free parameter that also enters as an argument of the strong coupling constant $\alpha_{s}(p_{T0}^2 + \mu_{R}^2)$.
As a default set in our calculations here we use $p_{T0} = 1$ GeV. This parameter could, in principle, be fitted to total charm cross section measured experimentally or calculated in the NLO/NNLO collinear calculations. The same method was also applied recently in the context of $J\!/\!\psi$-meson production in the color-evaporation model \cite{Maciula:2018bex}.
   
\subsection{Unintegrated parton distribution functions} 

\subsubsection{The CCFM uPDFs}

The CCFM evolution equation for gluon, in the limits of high and low energies (small- and large-$x$ values),
is almost equivalent to the BFKL and very similar to the DGLAP evolution, respectively \cite{CCFM}.
In order to correctly treat gluon coherence effects it introduces the so-called angular-ordering which is commonly considered as a great advantage of this framework.

In the leading logarthmic approximation, the CCFM equation 
for unintegrated gluon density ${\cal F}_{g}(x,k_t^2,\mu^2)$
can be written as
\begin{equation}
  \displaystyle {\cal F}_{g}(x,k_t^2,\mu^2) = {\cal F}_{g}^{(0)}(x,k_t^2,\mu_0^2) \Delta_s(\mu,\mu_0) + \atop { 
  \displaystyle + \int\frac{dz}{z}\int\frac{dq^2}{q^2}\Theta(\mu-zq)\Delta_s(\mu,zq) \tilde P_{gg}(z,k_t^2, q^2) {\cal F}_{g}\left(\frac{x}{z},k^{\prime \, 2}_t,q^2\right) },
\end{equation}

\noindent
where $\mu^2$ is the evolution (factorization) scale which is further defined by the maximum allowed
angle for any gluon emission, $k_t^\prime = q(1 - z) + k_T$
and $\tilde P_{gg}(z,k^2_t,q^2)$ is the CCFM
splitting function:
\begin{equation}
  \displaystyle \tilde P_{gg}(z,k^2_t,q^2) = \bar\alpha_s(q^2(1-z)^2) \left[\frac{1}{1-z}+\frac{z(1-z)}{2}\right] + \atop {
  \displaystyle + \bar\alpha_s(k_t^2)\left[\frac{1}{z}-1+\frac{z(1-z)}{2}\right]\Delta_{ns}(z,k^2_t,q^2) }.
\end{equation}

\noindent
The Sudakov and non-Sudakov form factors read:
\begin{equation}
 \ln \Delta_s(\mu,\mu_0)= - \int\limits_{\mu_0^2}^{\mu^2}\frac{d\mu^{\prime \, 2}}{\mu^{\prime \, 2}}\int\limits_0^{z_M=1-\mu_0/\mu^\prime}dz\,\frac{\bar\alpha_s(\mu^{\prime \, 2}(1-z)^2)}{1-z},
\end{equation}
\begin{equation}
\ln \Delta_{ns}(z,k_t^2, q_t^2) = -\bar\alpha_s(k_t^2)\int\limits_0^1\frac{dz^\prime}{z^\prime}\int\frac{dq^2}{q^2}\Theta(k_t^2-q^2)\Theta(q^2-z^{\prime\,2} q^2_t).
\end{equation}

\noindent
where $\bar \alpha_s = 3 \alpha_s/\pi$. 

The first term in the CCFM equation is the initial 
unintegrated gluon density multiplied by the Sudakov form factor.
It corresponds to the contribution of non-resolvable branchings between 
the starting scale $\mu_0^2$ and scale $\mu^2$.
The second term describes the details of the QCD evolution 
expressed by the convolution of the CCFM gluon splitting function
with the gluon density and the Sudakov form factor. The 
theta function introduces the angular ordering condition.
The CCFM equation can be solved numerically using 
the \textsc{updfevolv} program \cite{Hautmann:2014uua},
and the uPDFs for gluon and valence quarks can be 
obtained for any $x$, $k_t^2$ and $\mu^2$ values.

Within the CCFM approach the parton transverse momentum is allowed to be larger than the scale $\mu^2$.
This useful feature translates into the easiness of effective taking into account of higher-order radiative corrections,
that correspond to the initial-state real gluon
emissions which are resummed into the uPDFs. Thus, for any phenomenological studies in the $k_{T}$-factorization approach
the standard scheme with leading-order matrix elements is recommended as long as the CCFM uPDFs are used.   

\subsubsection{The PB uPDFs}

The Parton Branching (PB) method, introduced in Refs.~\cite{Hautmann:2017fcj,Martinez:2018jxt}, provides an iterative solution for the evolution of both collinear and transverse momentum dependent parton distributions. Within this novel method the splitting kinematics at each branching vertex stays under full control during the QCD evolution.
Here, soft-gluon emission in the region $z\to 1$ and transverse momentum recoils in the parton branchings along the QCD cascade are taken into account simultaneously. Therefore the PB approach allows for a natural determination of the uPDFs, as the transverse momentum at every branching vertex is known. It agrees with the usual methods to solve the DGLAP equations, but provides in addition a possibility to apply angular ordering instead of the standard ordering in virtuality.

Within the PB method, a soft-gluon resolution scale parameter $z_M$ is introduced into the QCD evolution equations that distinguish between non-resolvable and resolvable emissions. These two types of emissions are further treated with the help of the Sudakov form factors 
\begin{equation}
\label{sud-def}
 \Delta_a ( z_M, \mu^2 , \mu^2_0 ) = 
\exp \left(  -  \sum_b  
\int^{\mu^2}_{\mu^2_0} 
{{d \mu^{\prime 2} } 
\over \mu^{\prime 2} } 
 \int_0^{z_M} dz \  z 
\ P_{ba}^{(R)}\left(\alphas , 
 z \right) 
\right) 
  \;\; ,   
\end{equation}
and with the help of resolvable splitting probabilities $P_{ba}^{(R)} (\alphas,z)$, respectively.
Here $a , b$ are flavor indices, 
$\alphas$ is the strong coupling at a scale being a function of ${\mu}^{\prime 2}$,   
$z$ is 
the longitudinal momentum 
splitting variable, and   
$z_M < 1 $ is the soft-gluon resolution parameter.
Then, by connecting the evolution variable $\mu$ in the splitting process $b \to a c$
with the angle $\Theta$ of the momentum of particle $c$ with respect to the beam direction, 
the known angular ordering relation $\mu = | q_{t,c} | / (1 - z)$ is obtained, that ensures quantum coherence of softly radiated partons.

The \PBM\ evolution  equations with angular ordering condition for unintegrated parton densities    
$ {\cal F}_a ( x , k_{t} , \mu^2) $ 
are given 
by~\cite{Hautmann:2017fcj}   
\begin{eqnarray}
\label{evoleqforA}
   { {\cal F}}_a(x,k_t, \mu^2) 
 &=&  
 \Delta_a (  \mu^2  ) \ 
 { {\cal F}}_a(x,k_t,\mu^2_0)  
 + \sum_b 
\int
{{d^2 q_{t}^{\prime } } 
\over {\pi q_{t}^{\prime 2} } }
 \ 
{
{\Delta_a (  \mu^2  )} 
 \over 
{\Delta_a (  q_{t}^{\prime 2}  
 ) }
}
\ \Theta(\mu^2-q_{t}^{\prime 2}) \  
\Theta(q_{t}^{\prime 2} - \mu^2_0)
 \nonumber\\ 
&\times&  
\int_x^{z_M} {{dz}\over z} \;
P_{ab}^{(R)} (\alphas 
,z) 
\;{ {\cal F}}_b\left({x \over z}, k_{t}+(1-z) q_{t}^\prime , 
q_{t}^{\prime 2}\right)  
  \;\;  .     
\end{eqnarray}
Here, the starting disitribution for the uPDF evolution is taken in the factorized form as a product of collinear PDF fitted to the precise DIS data and an intrinsic transverse momentum distribution in a simple gaussian form. Unlike the CCFM parton distributions, the PB densities have the strong normalization property:
\be
\int { {\cal F}}_a(x,k_t, \mu^2)\; d k_t = f_{a}(x,\mu^2). 
\ee
The PB uPDFs can be calculated by an iterative Monte-Carlo method and are characterized by a steep drop of the parton densities at $k_{t}^2 > \mu^2$, again in contrast to the CCFM unintegrated distributions\footnote{Very recently, only a first attempt to incorporate CCFM effects into the PB method has been done \cite{Monfared:2019uaj}.}.
Therefore, for phenomenological studies within this model of unintegrated density, a higher-order scheme of the calculations is required that could compensate lack of extra emissions encoded in the uPDF. 

There are two available sets of the parton-branching uPDFs - PB-NLO-2018-set1 and PB-NLO-2018-set2, that correspond to different choice of the parameters of the initial distributions \cite{Martinez:2018jxt}. Both of them, including uncertainties are available in TMDLIB \cite{Hautmann:2014kza}. In the following, the PB-NLO-2018-set1 uPDFs were used in numerical calculations.

\subsubsection{The KMR/MRW uPDFs}

Another DGLAP-based and frequently used in phenomenological studies prescription for unintegrated gluon densities is the Kimber-Martin-Ryskin (KMR) approach \cite{Kimber:1999xc,Kimber:2001sc,Watt:2003mx}. It has been successfully used especially for charm production at the LHC, including inclusive charm \cite{Maciula:2013wg}, charm-anticharm pairs \cite{Maciula:2013wg,Karpishkov:2016hnx}, double and triple charm \cite{Maciula:2013kd,Maciula:2017meb}, as well as charm associated with jets \cite{Maciula:2017egq}. According to this approach the unintegrated gluon distribution is given
by the following formula
\begin{eqnarray} \label{eq:UPDF}
  f_g(x,k_t^2,\mu^2) &\equiv& \frac{\partial}{\partial \log k_t^2}\left[\,g(x,k_t^2)\,T_g(k_t^2,\mu^2)\,\right]\nonumber \\ &=& T_g(k_t^2,\mu^2)\,\frac{\alpha_S(k_t^2)}{2\pi}\,\sum_{b }\,\int_x^1\! d z\,P_{gb}(z)\,b \left (\frac{x}{z}, k_t^2 \right).
\end{eqnarray}
This formula makes sense for $k_t > \mu_0$, where $\mu_0\sim 1$ GeV
is the minimum scale for which DGLAP evolution of the conventional
collinear gluon PDF, $g(x,\mu^2)$, is valid.

The virtual (loop) contributions may be resummed to all orders by the Sudakov form factor
\begin{equation} \label{eq:Sudakov}
  T_g (k_t^2,\mu^2) \equiv \exp \left (-\int_{k_t^2}^{\mu^2}\!\diff{\kappa_t^2}\,\frac{\alpha_S(\kappa_t^2)}{2\pi}\,\sum_{b}\,\int_0^1\!\dif{z}\;z \,P_{b g}(z) \right ),
\end{equation}
which gives the probability of evolving from a scale $k_t$ to a scale $\mu$ without parton emission.
The exponent of the gluon Sudakov form factor can be simplified using the following identity: $P_{qg}(1-z)=P_{qg}(z)$.  Then the gluon Sudakov form factor reads
\begin{equation}
  T_g(k_t^2,\mu^2) = \exp\left(-\int_{k_t^2}^{\mu^2}\!\diff{\kappa_t^2}\,\frac{\alpha_S(\kappa_t^2)}{2\pi}\,\left( \int_{0}^{1-\Delta}\!\dif{z}\;z \,P_{gg}(z) + n_F\,\int_0^1\!\dif{z}\,P_{qg}(z)\right)\right),
\end{equation}
where $n_F$ is the quark--antiquark active number of flavours into which the gluon may split. Due to the presence of the Sudakov form factor in the KMR prescription only last emission generates transverse momentum of the gluons initiating hard scattering.

In the above equation the variable $\Delta$ introduces a restriction of the phase space for gluon emission and is crucial for the final shape and characteristics of the unintegrated density. In Ref.~\cite{Kimber:1999xc} the cutoff $\Delta$ was set in accordance with the strong ordering (SO) in transverse momenta of the real parton emission in the DGLAP evolution, 
\be
\label{eq:11}
\Delta =\frac{k_{t}}{\mu}\,.
\ee
This corresponds to the orginal KMR prescription where one always has $k_{t}^{2} < \mu_{F}^{2}$ restriction and the Sudakov form-factor always satisfies the $T_{g}(k_{t}^{2},\mu^{2})<1$ condition.

The prescription for the cutoff $\Delta$ was further modified in Ref.~\cite{Kimber:2001sc,Watt:2003mx} to account for the angular ordering 
(AO) in parton emissions in the spirit of the CCFM evolution,
\be
\label{eq:13}
\Delta=\frac{k_{t}}{k_{t}+\mu}\,.
\ee
This modification leads to a bigger upper limit for $k_{t}$ than in the DGLAP scheme and opens the $k_{t}^{2} > \mu_{F}^{2}$ region.
In this extra kinematical regime one gets $T_{g}(k_{t}^{2},\mu^{2})>1$, which
contradicts its interpretation as a probability of no real emission. Thus, the Sudakov form factor is there usually set to be equal one.
For transparency, here the modified KMR model will be referred to as the Martin-Ryskin-Watt (MRW) model \cite{Watt:2003mx}.

Different definitions of the ordering cut-off lead to significant differences between the two models. In the KMR model the $k_{t}^{2} > \mu_{F}^{2}$ region is forbidden while in the MRW case the $k_{t}^{2} > \mu_{F}^{2}$ contributions are directly allowed (see \textit{e.g.} a detailed discussion in Ref.~\cite{Golec-Biernat:2018hqo}). In the MRW model both in quark and gluon densities large $k_{t}$-tails appear, in contrast to the KMR case.
These two models need to be therefore differently treated in phenomenological applications. The MRW model shall be used with the standard $k_{T}$-factorization scheme at leading-order (as in the CCFM uPDFs case), while the original KMR model requires the new procedure with the leading-order mechanisms matched with higher-order contributions (as in the PB uPDFs case). 

In the numerical calculations below we used the CT14lo collinear PDFs \cite{Dulat:2015mca} to calculate both, the KMR and the MRW unintegrated densities. 

\subsection{Open charm meson production}

The transition of charm quarks to open charm mesons is done in the framework of the independent parton fragmentation picture (see \textit{e.g.} Refs.~\cite{Maciula:2015kea,Maciula:2019iak}).
Here we follow the standard prescription, where the inclusive distributions of open charm meson are obtained through a convolution of inclusive distributions of charm quarks/antiquarks and $c \to D$ fragmentation functions:
\begin{equation}
\frac{d \sigma(pp \rightarrow D X)}{d y_D d^2 p_{t,D}} \approx
\int_0^1 \frac{dz}{z^2} D_{c \to D}(z)
\frac{d \sigma(pp \rightarrow c X)}{d y_c d^2 p_{t,c}}
\Bigg\vert_{y_c = y_D \atop p_{t,c} = p_{t,D}/z} \;,
\label{Q_to_h}
\end{equation}
where $p_{t,c} = \frac{p_{t,D}}{z}$ and $z$ is the fraction of
longitudinal momentum of charm quark $c$ carried by a meson $D$.
In the numerical calculations we take the Peterson fragmentation function \cite{Peterson:1982ak}, often used in the context of hadronization of heavy flavours. 
Then, the hadronic cross section is normalized by the relevant charm fragmentation fractions for a given type of $D$ meson \cite{Lisovyi:2015uqa}.  
In the numerical calculations below for $c \to D^{0}$ meson transition we take the fragmentation probability $\mathrm{P}_{c \to D} = 61\%$.

\section{Numerical results}

In this section we present our numerical results for $D^{0} + \overline{D^{0}}$ meson production in fixed-target $p+^4\!\mathrm{He}$ collisions at $\sqrt{s_{NN}}=86.6$ GeV, 
measured for the first time very recently by the LHCb collaboration \cite{Aaij:2018ogq}.
The measured cross section for $D^{0} + \overline{D^{0}}$ final state in the LHCb acceptance is
\be
\sigma_{\mathrm{LHCb}} = 80.8 \pm 2.2(\mathrm{stat}) \pm 6.3(\mathrm{syst}) \; \mu b/\mathrm{nucleon} . 
\ee   
The experimental cross sections are divided by the number of nucleons and are compared below with the theoretical results for $pp$-scattering. Nuclear effects in the case of the
$p+^4\!\mathrm{He}$ interactions and for considered kinematical range are expected to be negligible, which was checked and explicitly shown in Ref.~\cite{Aaij:2018ogq} (see Fig.4 therein), and therefore are also neglected here.
Below we show in addition the theory/data ratio.

As a default set in the numerical calculations we take the renormalization/factorization scales
$\mu^2 = \mu_{R}^{2} = \mu_{F}^{2} = \sum_{i=1}^{n} \frac{m^{2}_{it}}{n}$ (averaged tranvserse mass of the given final state) and the charm quark mass $m_{c}=1.5$ GeV. The strong-coupling constant $\alpha_{s}(\mu_{R}^{2})$ at leading-order is taken from the CT14 PDF routines.

\subsection{The $k_{T}$-factorization scheme with the CCFM uPDFs} 

\begin{figure}[!h]
\begin{minipage}{0.35\textwidth}
  \centerline{\includegraphics[width=1.0\textwidth]{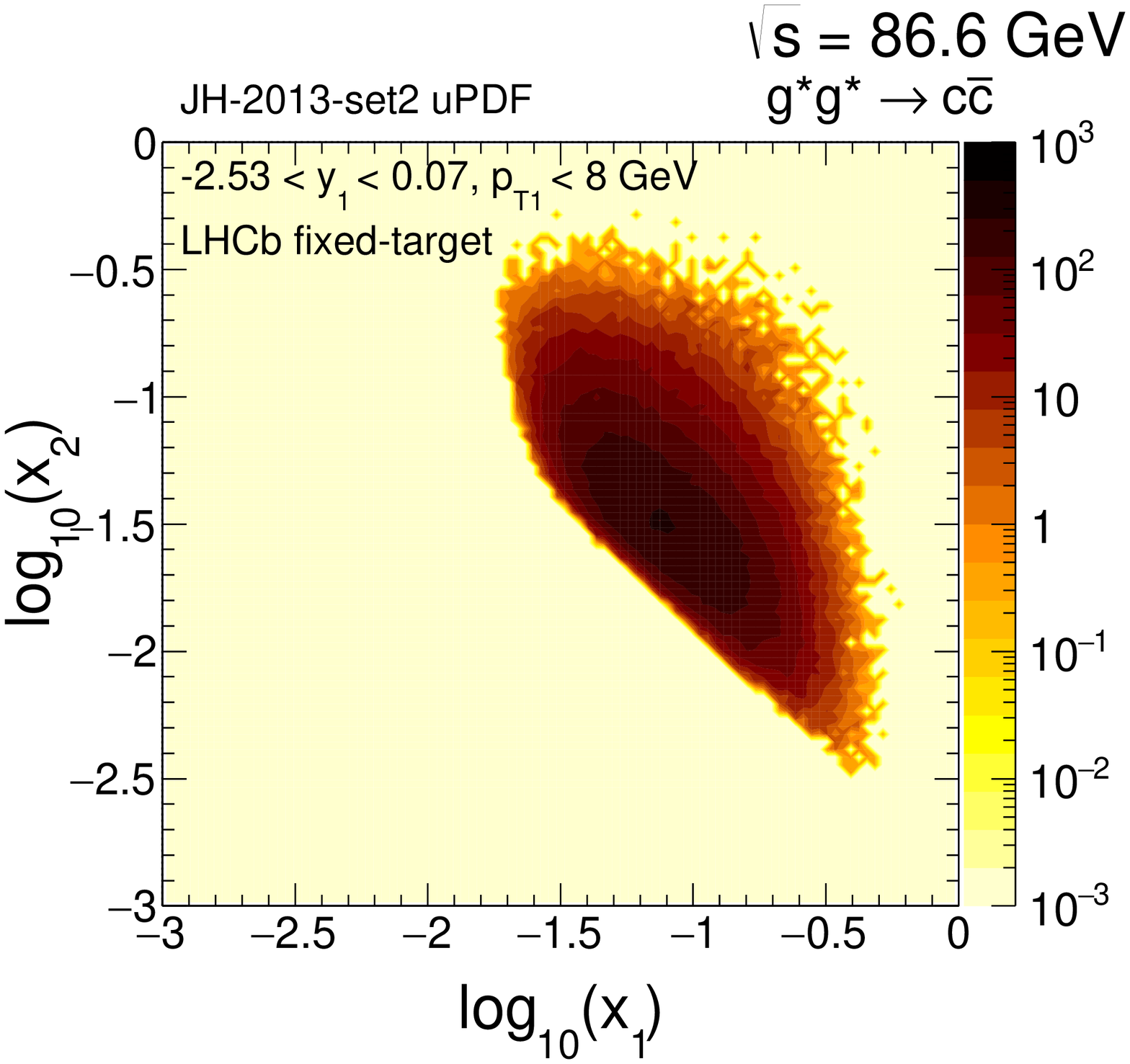}}
\end{minipage}
\hspace{1cm}
\begin{minipage}{0.35\textwidth}
  \centerline{\includegraphics[width=1.0\textwidth]{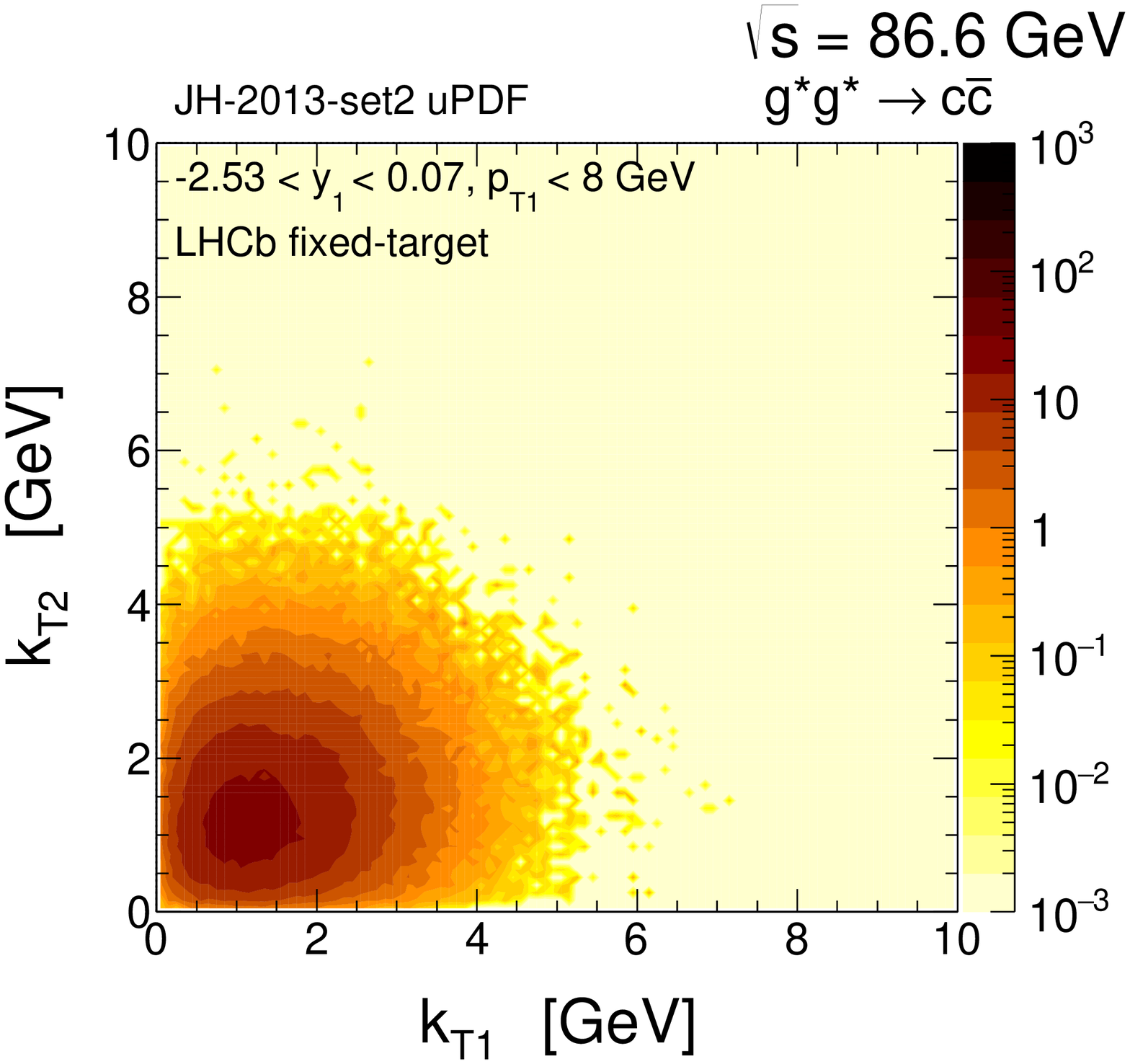}}
\end{minipage}
  \caption{
\small The two-dimensional differential cross sections for $c\bar c$-pair production in proton-proton scattering at $\sqrt{s}=86.6$ GeV in microbarns as a function of longitudinal momentum fractions $\log_{10}(x_1)$ and $\log_{10}(x_2)$ (left panel) and transverse momenta $k_{t1}$ and $k_{t2}$ (right panel) of the incident gluons. Other details are specified in the figure.   
}
\label{fig:2dim}
\end{figure}

We start with the results for the $g^*g^* \to c\bar c$ mechanism calculated in the framweork of the $k_{T}$-factorization approach with off-shell initial state partons and with the CCFM uPDFs. Before we go to the main results for the $D^{0}$ meson cross sections and their comparison with the LHCb fixed-target data we wish to present complementary plots that will be helpful in qualitative visualization of the kinematics behind the considered production mechanism. In Fig.~\ref{fig:2dim} we present double differential parton-level cross section for charm quarks as a function of
longitudinal momentum fractions $\log_{10}(x_1)$ and $\log_{10}(x_2)$ (left panel) and transverse momenta $k_{t1}$ and $k_{t2}$ (right panel) of the incident gluons. Here we impose the cuts relevant for the LHCb fixed-target mode on one of the quarks from the $c\bar c$-pair. We clearly see that within the present phenomenological analysis we probe the unintegrated gluon distributions at large $x$-values with maximum of the cross section around $10^{-1}$. The transverse momenta of the initial state gluons are quite small here ($< 5$ GeV) with maximum of the cross section between 1 and 2 GeV which is at the border of nonperturbative and perturbative regions. This so far unexplored kinematical domain is of course very interesting and could help to constrain gluon uPDFs in these exotic limits.      
 
\begin{figure}[!h]
\begin{minipage}{0.47\textwidth}
  \centerline{\includegraphics[width=1.0\textwidth]{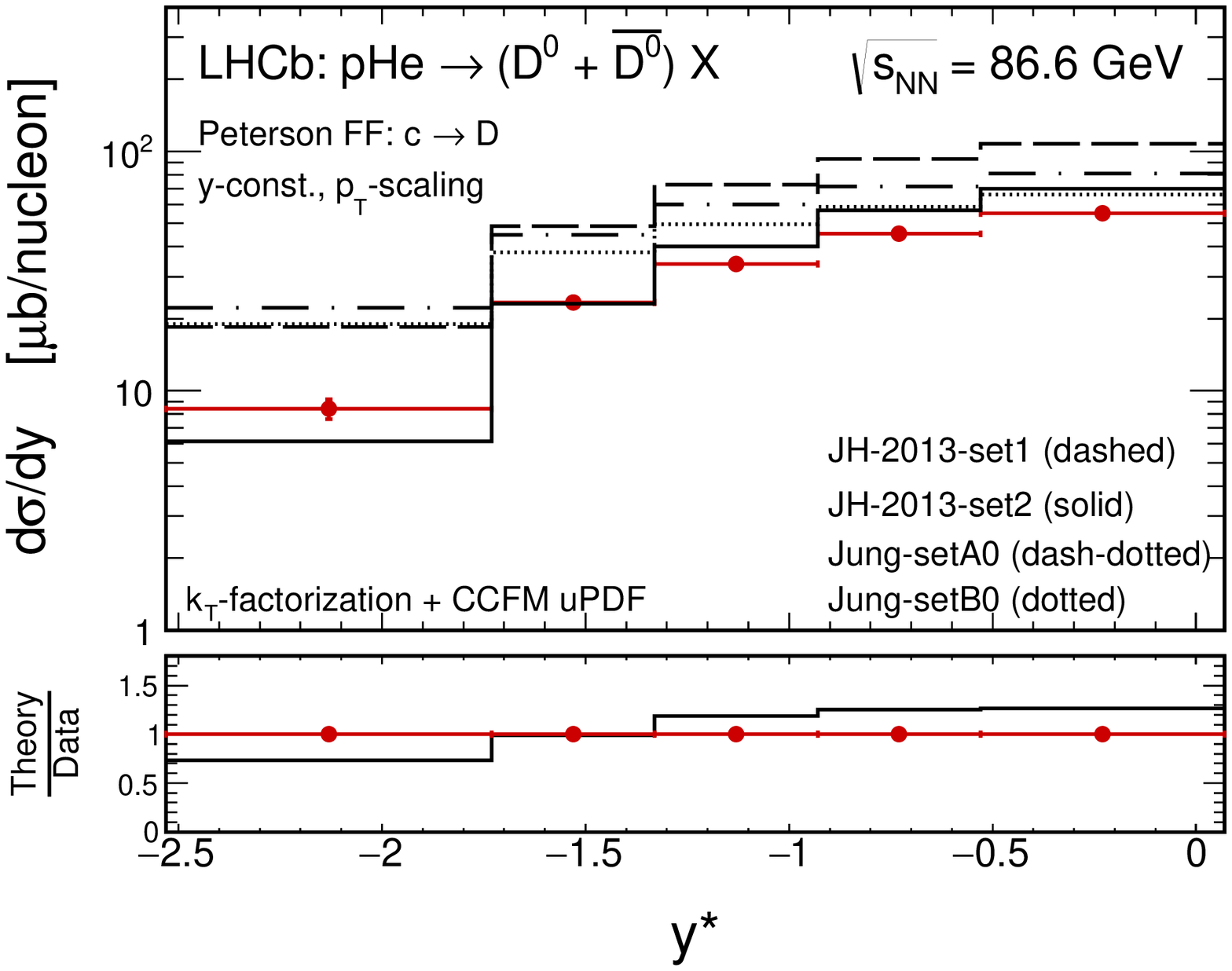}}
\end{minipage}
\begin{minipage}{0.47\textwidth}
  \centerline{\includegraphics[width=1.0\textwidth]{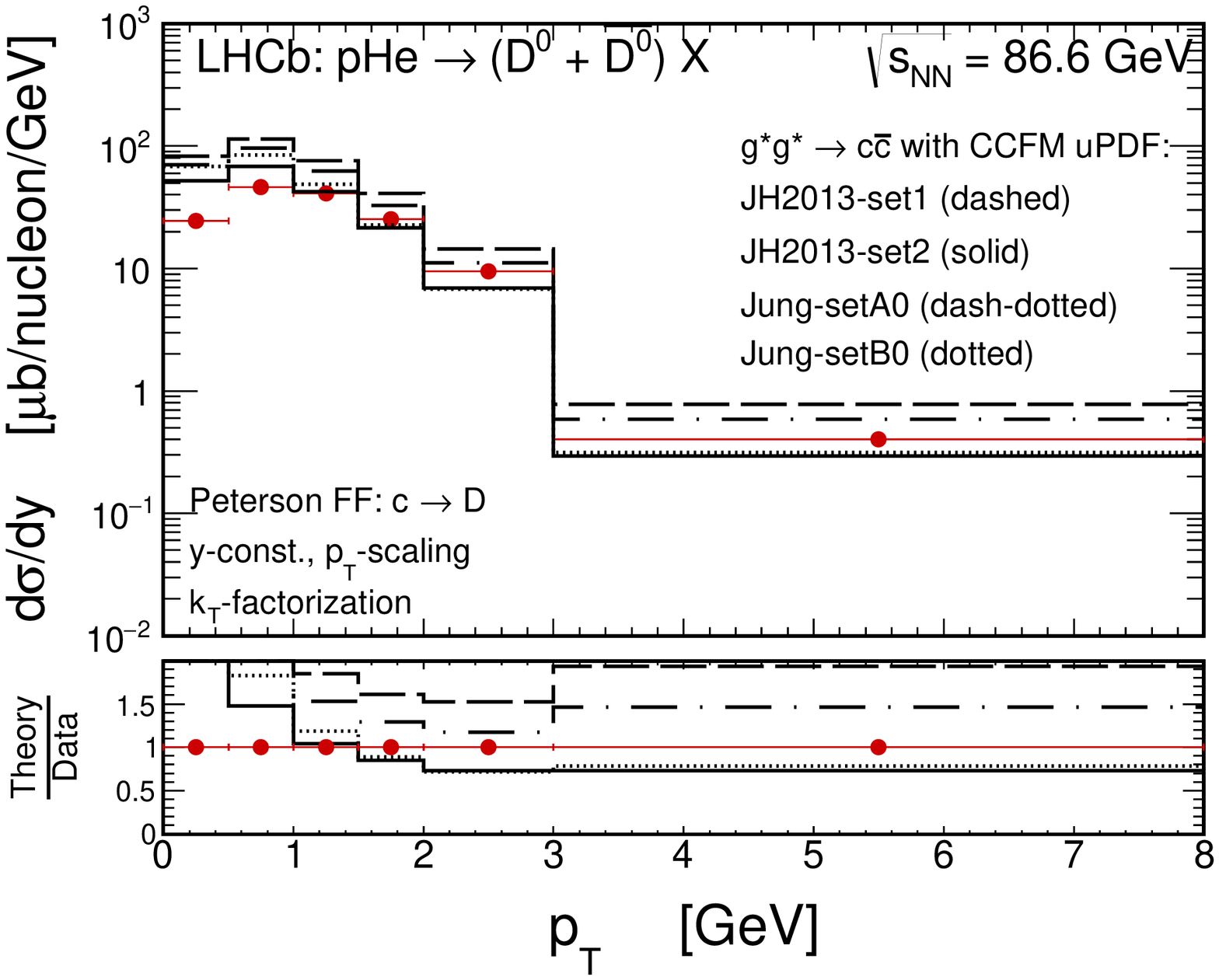}}
\end{minipage}
  \caption{
\small Rapidity (left) and transverse momentum (right) distributions of $D^{0}$ meson (plus $\overline{D^{0}}$ antimeson)
for $p+^4\!\mathrm{He}$ collisions together with the LHCb data. Here results of the $k_{T}$-factorization calculations for four different gluon CCFM uPDFs are shown. Other details are specified in the figure.   
}
\label{fig:CCFM}
\end{figure}

In Fig.~\ref{fig:CCFM} we present differential distributions of $D^{0}$ meson as a function of c.m.s. rapidity (left panel) and transverse momentum (right panel) for $p+^4\!\mathrm{He}$ collisions together with the LHCb experimental data points \cite{Aaij:2018ogq}.
For the numerical predictions presented here we have used four different sets of the gluon CCFM uPDFs: the most up-to-date JH-2013-set1 and JH-2013-set2 fits \cite{Hautmann:2013tba}, as well as a bit older Jung-setA0 and Jung-setB0 \cite{Jung:2004gs}. Here, in the CCFM scheme we use an atypical value for factorization scale $\mu_{F}^2 = M_{c\bar c}^2 + P_{T}^{2}  $, where $M_{c\bar c}$ and $P_{T}$ are the $c\bar c$-invariant mass (or energy of the scattering subprocess) and the transverse momentum of $c\bar c$-pair (or the incoming off-shell gluon pair). This rather unusual definition has to be applied as a consequence of the CCFM evolution algorithm \cite{Hautmann:2013tba}. From the comparison of our numerical results with the LHCb data we conclude that the CCFM scheme of the calculations with the JH-2013--set2 gluon uPDF leads to a very good description of the experimental data points. Rest of the used CCFM uPDFs seems to slightly overestimate the LHCb data. Both, the JH-2013-set1 and JH-2013-set2 gluon densities are determined from high-precision DIS measurements, including experimental and theoretical uncertainties.
However, the JH-2013-set1 is determined from the fit to inclusive $F_2$ data only while the JH-2013-set2 is determined from the fit to both $F^{(\mathrm{charm})}_2$ and $F_2$ data. We see that in our calculations here these two sets lead to quite different results. The JH-2013-set1 gluon uPDF results in a larger charm cross section and overestimates the experimental data. The LHCb fixed-target open charm data visibly prefers the JH-2013-set2 gluon uPDF. The integrated cross section for $D^{0} + \overline{D^{0}}$ final state in the LHCb kinematics obtained with the JH-2013-set2 gluon uPDF is $\sigma_{\mathrm{CCFM}} = 102.15 \; \mu b $.

\begin{figure}[!h]
\begin{minipage}{0.47\textwidth}
  \centerline{\includegraphics[width=1.0\textwidth]{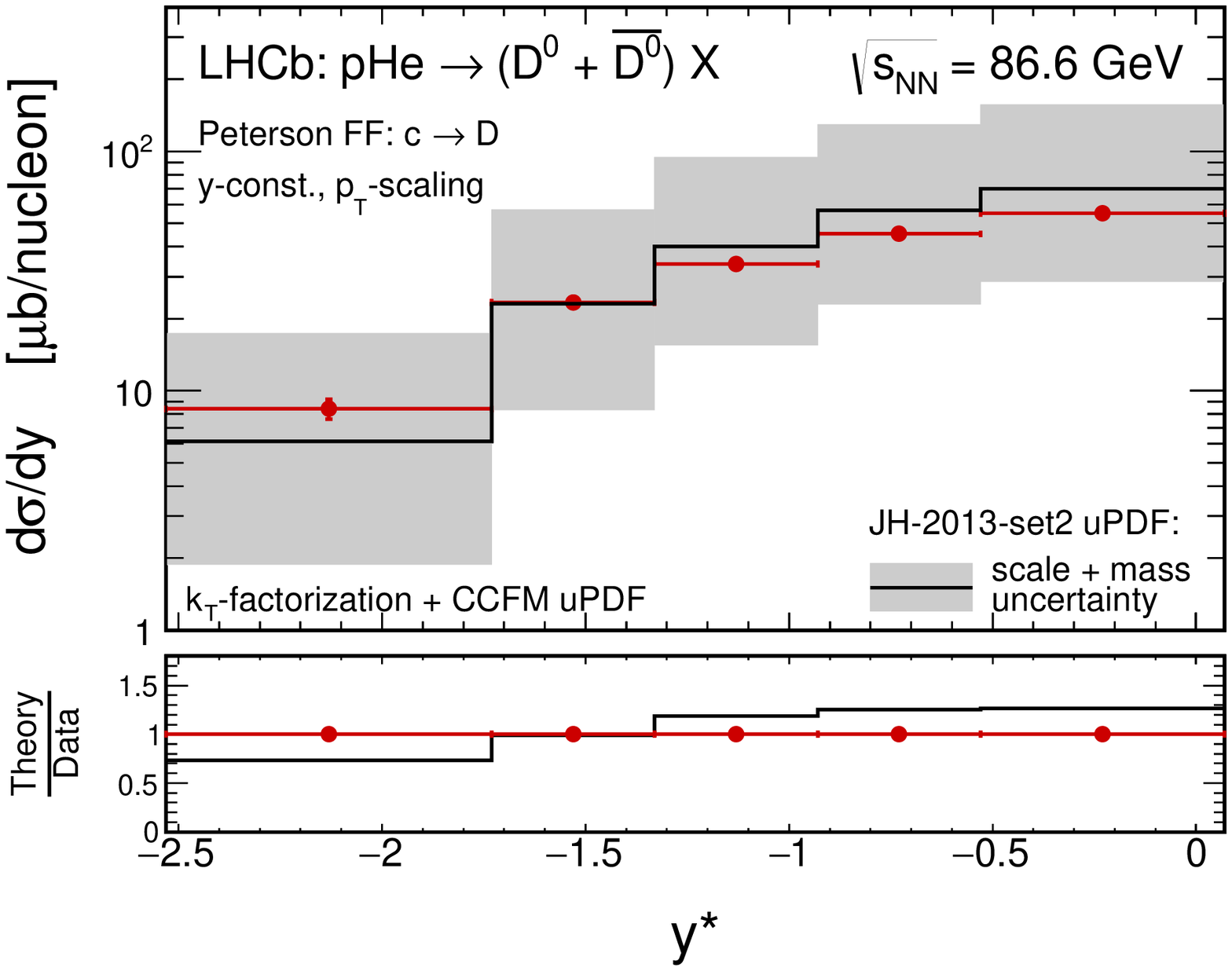}}
\end{minipage}
\begin{minipage}{0.47\textwidth}
  \centerline{\includegraphics[width=1.0\textwidth]{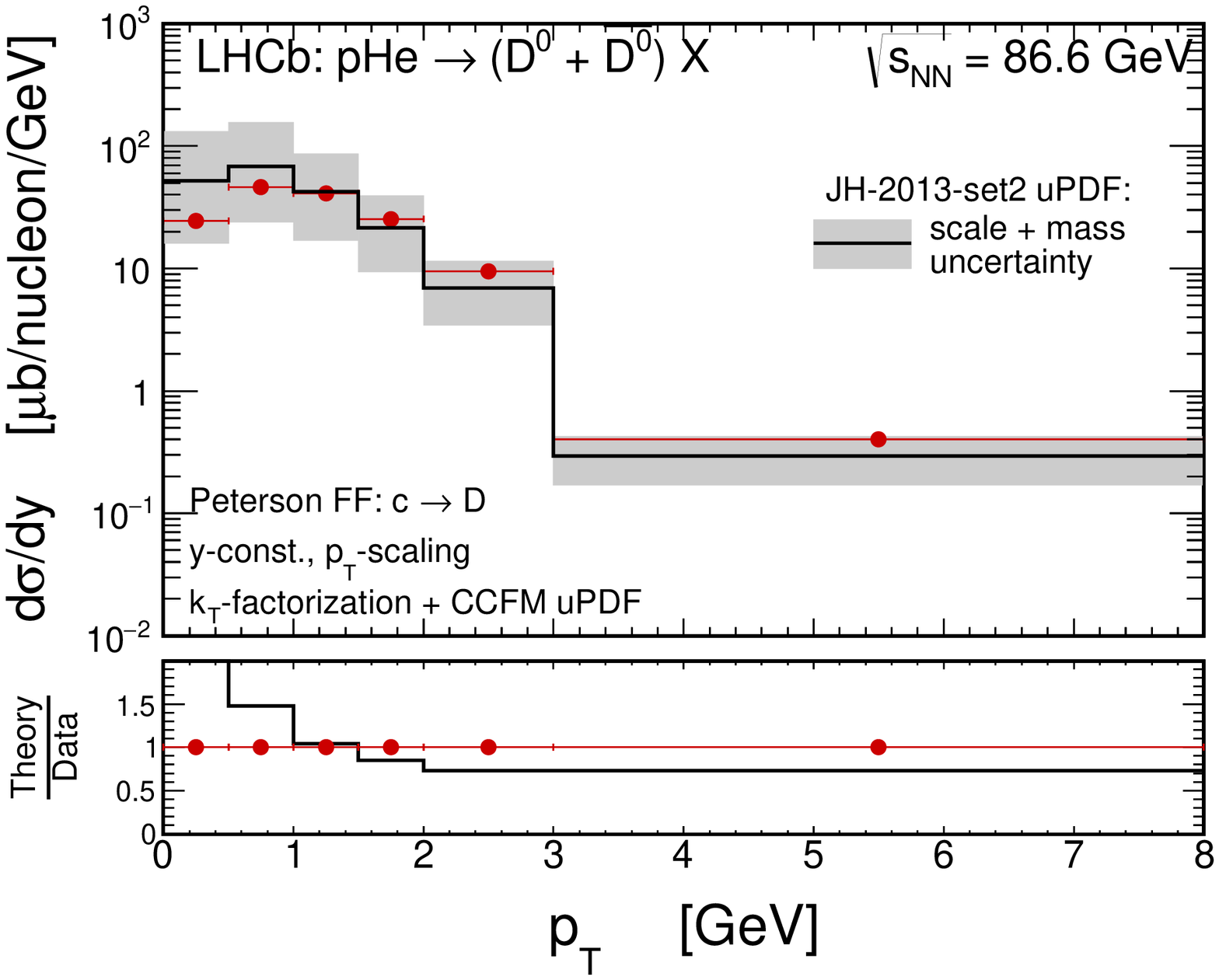}}
\end{minipage}
  \caption{
\small Rapidity (left) and transverse momentum (right) distributions of $D^{0}$ meson (plus $\overline{D^{0}}$ antimeson)
for $p+^4\!\mathrm{He}$ collisions together with the LHCb data. Here results of the $k_{T}$-factorization calculations for the JH-2013-set2 gluon CCFM uPDF are shown together with corresponding uncertainty bands. The shaded bands represent the renormalization/factorization scale and quark mass uncertainties summed in quadrature.
}
\label{fig:CCFM-uncert}
\end{figure}

In Fig.~\ref{fig:CCFM-uncert} we show in addition uncertainties of the calculations done for the JH-2013-set2 gluon uPDF. The shaded bands here correspond to uncertainties of the calculations related to the renormalization/factorization scales and with the charm quark mass summed in quadrature. The scales $\mu^2=\mu_{R}^2=\mu_{F}^2$ are divided or multiplied by a factor of 2 with respect to the central value and similarly charm quark mass is varied as follows: $m_{c} = 1.5 \pm 0.25$ GeV. Taking into account the theoretical uncertainties we get an excellent description of the data in the whole considered kinematical regime. The uncertainties are quite large at very low transverse momenta of $D^{0}$ meson where the uncertainty of charm quark mass plays an important role. The large uncertainty of the first bin in transverse momentum affects the whole spectrum in meson rapidity. At larger meson $p_{T}$'s the overall uncertainty is $\lesssim 2$ which is rather standard in any pQCD calculations.

\subsection{The $k_{T}$-factorization scheme with Parton-Branching uPDF}

Above we have exactly shown that the $k_{T}$-factorization approach with the CCFM uPDFs works very well for charm production already at leading-order, when only leading order $g^*g^* \to c \bar c$ mechanism is taken into account. As was already discussed, in this framework, higher-order radiative corrections (namely, a part of NLO, and even NNLO terms corresponding to the initial-state real gluon emissions) are effectively taken into account. Technically, it is driven by the fact that in the CCFM scheme the incoming gluons are allowed to have transverse momenta larger than the factorization scale, i.e. $k_{t}^2 > \mu^2$. Ipso facto, having hard extra emissions from uPDFs one takes into account in an effective way \textit{e.g.} flavour excitation and gluon splitting mechanisms, which are recognized to be of a special importance in the case of heavy flavour production \cite{Norrbin:2000zc}.     

\begin{figure}[!h]
\begin{minipage}{0.47\textwidth}
  \centerline{\includegraphics[width=1.0\textwidth]{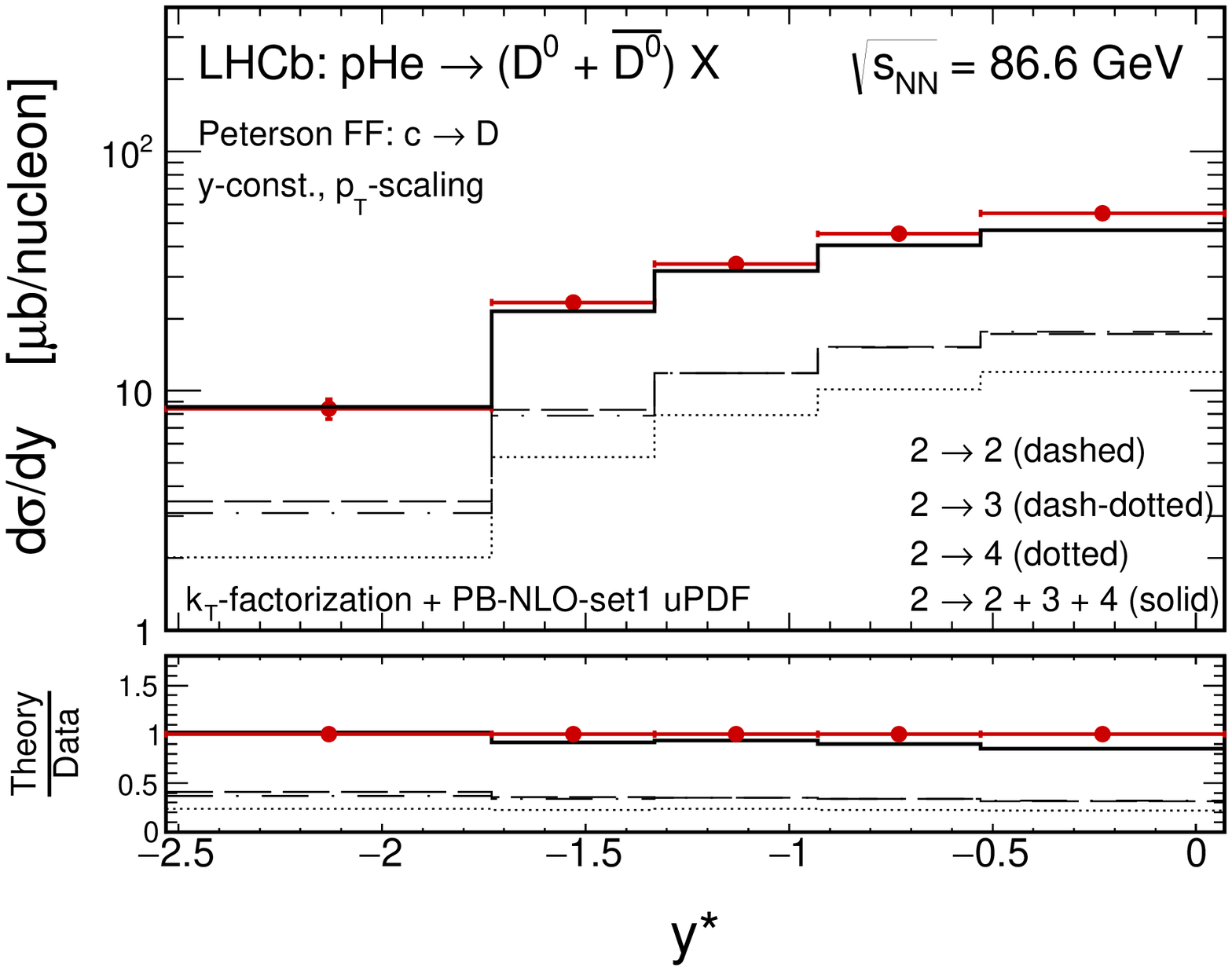}}
\end{minipage}
\begin{minipage}{0.47\textwidth}
  \centerline{\includegraphics[width=1.0\textwidth]{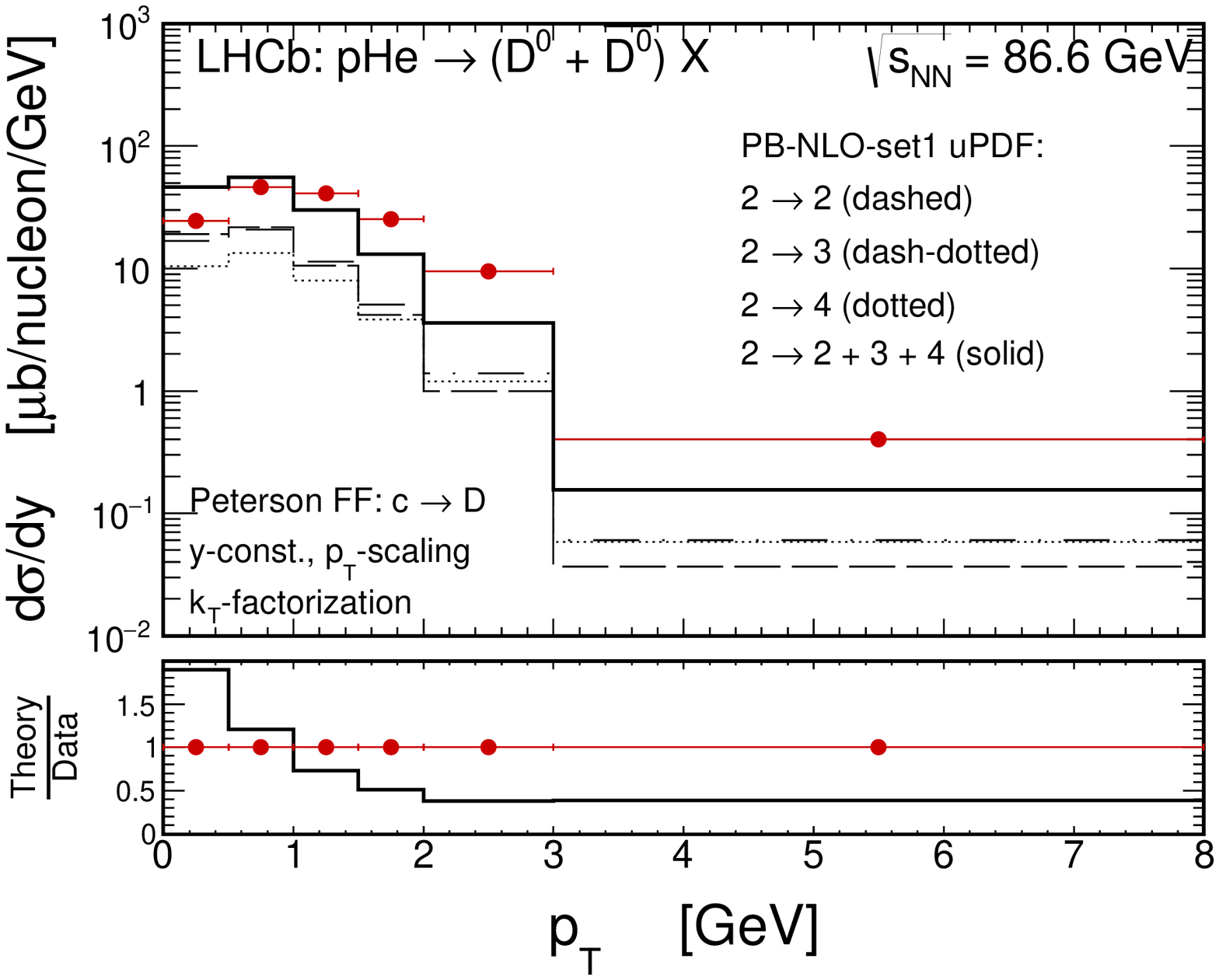}}
\end{minipage}
  \caption{
\small Rapidity (left) and transverse momentum (right) distributions of $D^{0}$ meson (plus $\overline{D^{0}}$ antimeson)
for $p+^4\!\mathrm{He}$ collisions together with the LHCb data. Here the $k_{T}$-factorization results with the PB-NLO-set1 uPDF for the $2\to 2$, $2\to 3$ and $2\to 4$ mechanisms are shown separately. Other details are specified in the figure.  
}
\label{fig:PB}
\end{figure}

As it was already mentioned, in contrast to the CCFM calculations, in the PB scheme which is based on full-flavour DGLAP evolution one has to include the usual leading order $g^*g^* \to c \bar c$ mechanism properly matched with a number of additional higher-order terms. Here, no extra hard emissions from uPDFs are allowed and they must be taken into account at the level of hard matrix elements. In our calculations here we take into account all gluon and quark-induced off-shell subprocesses of the $2 \to 3$ and the $2\to 4$ type at tree-level, corresponding to associated production of charm with one and two (mini)jets, respectively. To avoid a possible double-counting, the $2 \to 2$, $2 \to 3$ and $2 \to 4$ contributions are added together according to the matching procedure proposed very recently in Ref.~\cite{Maciula:2019izq} for the case of similar studies of charm production but at higher energy $\sqrt{s} = 7$ TeV.

In Fig.~\ref{fig:PB} we present results of the PB scheme of the calculations based on the $k_{T}$-factorization approach.
We use the PB-NLO-set1 quark and gluon uPDFs to calculate again the differential distributions of $D^{0}$ meson as a function of c.m.s. rapidity (left panel) and transverse momentum (right panel). The dashed, dash-dotted and dotted histograms correspond to the $2 \to 2$, $2 \to 3$ and $2 \to 4$ mechanisms, respectively. The solid histogram presents their sum denoted as $2 \to 2 + 3+ 4$. We see that in the PB scheme, as one could expect, the leading-order
$2 \to 2$ mechanism is completely insufficient and significantly underestimates the LHCb data points. Within the PB scheme
a reasonable description of the data can be achieved only when the leading-order $2 \to 2$ mechanism is supplemented by the higher-order $2 \to 3$ and $2 \to 4$ contributions. The $2 \to 2 + 3+ 4$ result stays in an excellent agreement with the measured rapidity distributions. The transverse momentum distribution is also well described at small transverse momenta, however, at larger $p_{T}$'s some small missing strength with respect to the data appears. The integrated cross section for $D^{0} + \overline{D^{0}}$ final state in the LHCb kinematics obtained with the PB-NLO-set1 gluon uPDF is $\sigma_{\mathrm{PB}} = 78.19 \; \mu b $.

Comparing the final result of the PB scheme with the PB-NLO-set1 uPDF and the result of the CCFM scheme with the JH2013-set2 uPDF we conclude that both prescriptions lead to a very similar and consistent results (up to factor 2) and provides a good quality description of the LHCb fixed-target charm data.  

\subsection{The $k_{T}$-factorization scheme with the KMR/MRW uPDF}

The difference between the KMR and the MRW model is similar to the case of the CCFM and the PB unintegrated densities discussed above. Thus, in the calculations with the MRW uPDF only the leading-order $g^*g^* \to c \bar c$ mechanism has to be taken into account. On the other hand, when applying the KMR uPDF one has to follow the calculation scheme with the $2 \to 2$, $2 \to 3$ and $2 \to 4$ tree-level subprocesses taken into account at the level of hard-matrix elements. It was shown in Ref.~\cite{Maciula:2019izq} that both schemes of the calculation are convergent and lead to very similar results in the case of charm production at $\sqrt{s} = 7$ TeV collision energy.

The $k_{T}$-factorization scheme of the calculation with the MRW uPDF was supported and justified by various open charm (and not only) LHC data sets. However, some criticism of the MRW model was expressed very recently \textit{e.g.} in Ref.~\cite{Guiot:2019vsm}. It is also commonly questioned, whether the MRW model could be successfully used to describe DIS high-precision data \cite{Kutak}. Fortunately, the recently proposed model of $k_{T}$-factorization calculations with the higher-order corrections at tree-level \cite{Maciula:2019izq} brings an useful alternative for the standard calculations and seems to be a good way to avoid the mentioned problems.

\begin{figure}[!h]
\begin{minipage}{0.47\textwidth}
  \centerline{\includegraphics[width=1.0\textwidth]{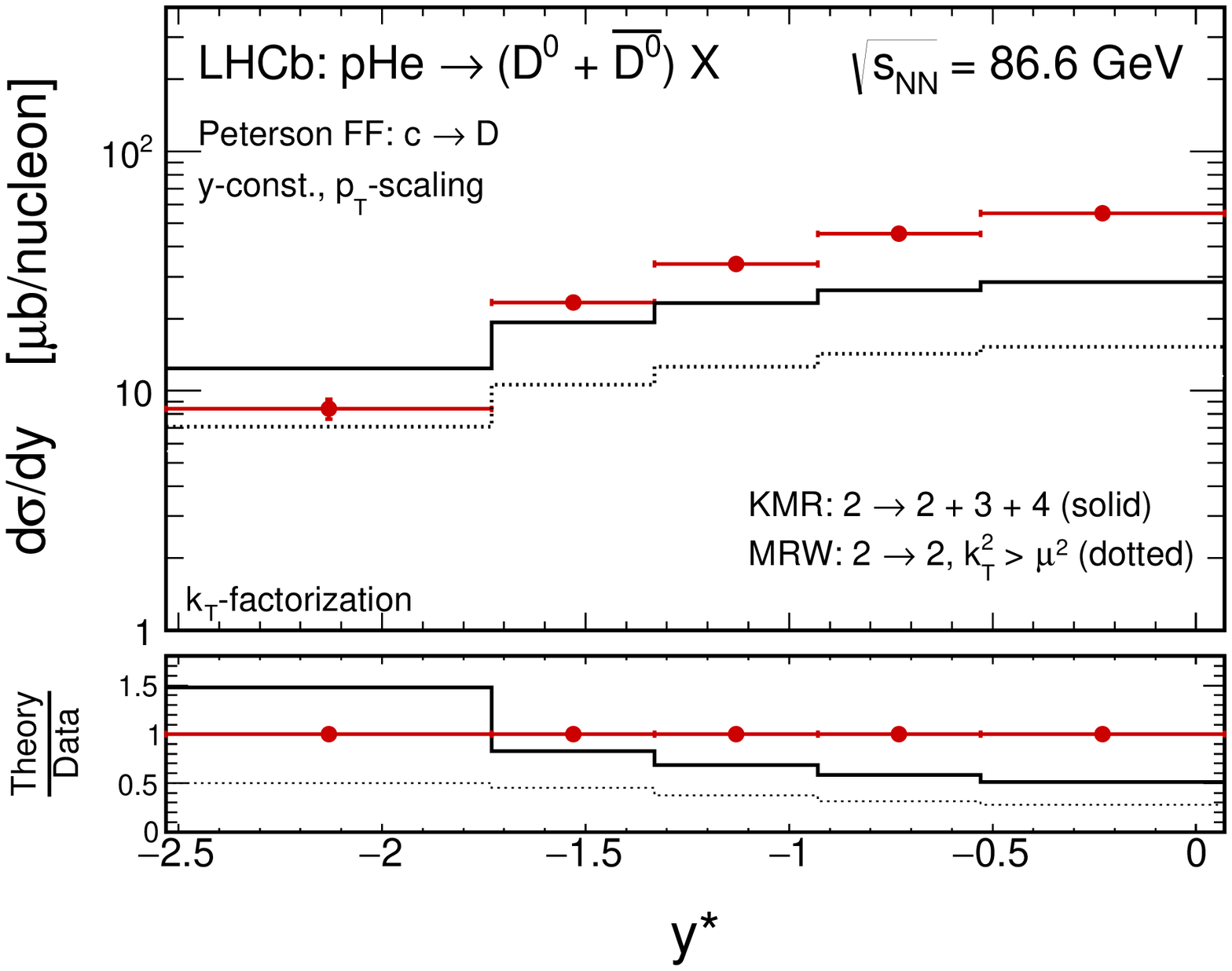}}
\end{minipage}
\begin{minipage}{0.47\textwidth}
  \centerline{\includegraphics[width=1.0\textwidth]{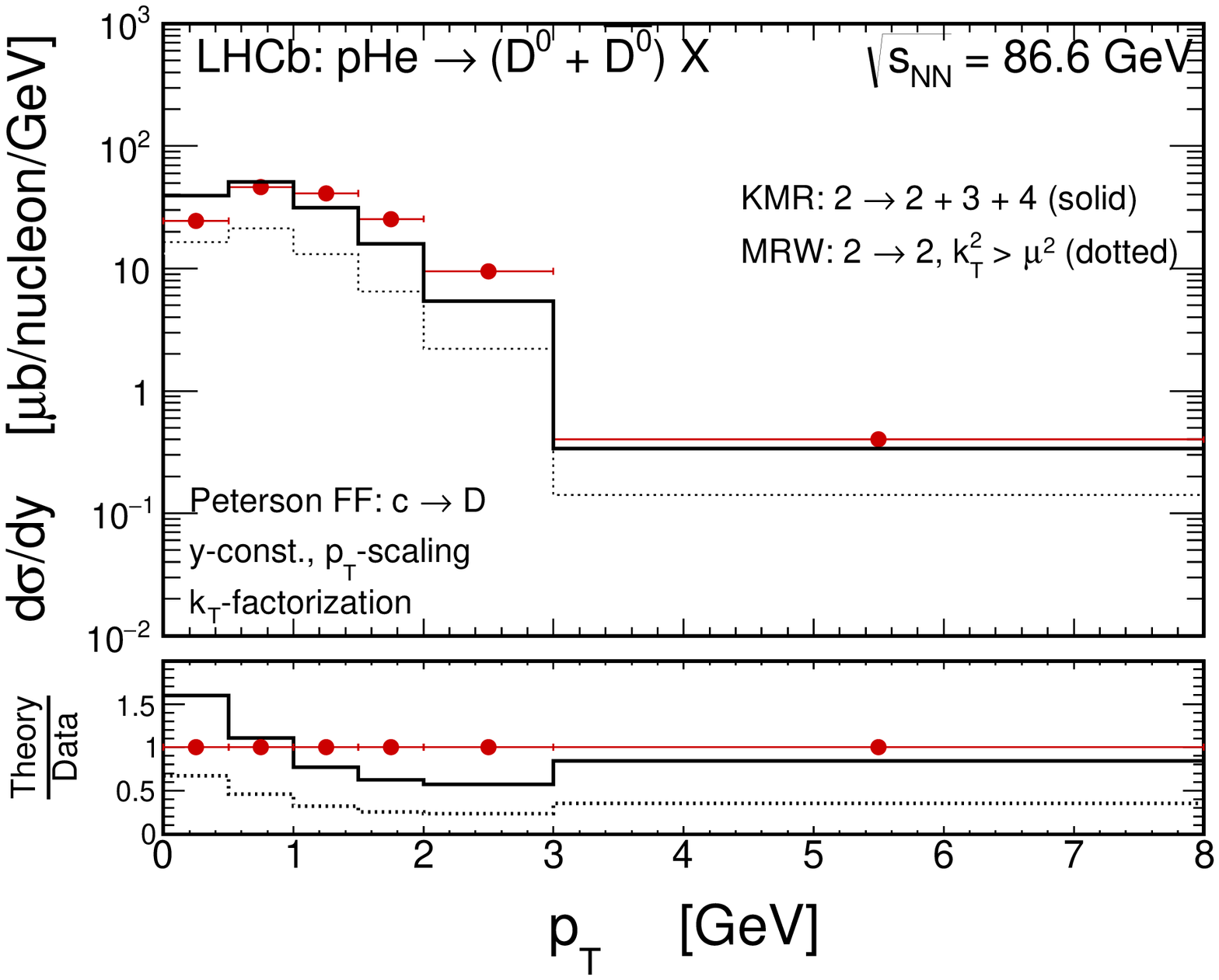}}
\end{minipage}
  \caption{
\small Rapidity (left) and transverse momentum (right) distributions of $D^{0}$ meson (plus $\overline{D^{0}}$ antimeson)
for $p+^4\!\mathrm{He}$ collisions together with the LHCb data. Here the $k_{T}$-factorization results with the KMR-CT14lo (solid) and with the MRW-CT14lo (dotted) uPDFs are compared. Other details are specified in the figure.  
}
\label{fig:KMR}
\end{figure}

Here, we wish to compare the results of the calculation with the MRW uPDF and the results of the calculation with the new $2 \to 2 + 3+ 4$ scheme       
relevant for the KMR uPDF. Such a comparison at $\sqrt{s}=86.6$ GeV may help to judge whether the agreement between the $\sqrt{s} = 7$ TeV LHC open charm data and the $k_T$-factorization predictions with the MRW uPDFs (previously obtained by different authors) is kind of accidental or not. In Fig.~\ref{fig:KMR} we show again the rapidity (left) and the transverse momentum (right) differential distributions of $D^{0}$ meson at $\sqrt{s}= 86.6$ GeV.
The solid histograms correspond to the results obtained with the KMR uPDFs while the dashed ones are for the calculations with the MRW uPDFs. We observe a large difference between the two results - unlike the case of the $\sqrt{s}=7$ TeV studies (see Fig.15 in Ref.~\cite{Maciula:2019izq}). The results obtained with the $2 \to 2 + 3+ 4$ scheme and the KMR uPDF lie much closer the data points. The quality of the data description is very similar as in the case of the CCFM uPDF. On the other hand, the results of the standard calculations with the MRW uPDF significantly underestimates the LHCb data here. The integrated cross sections for $D^{0} + \overline{D^{0}}$ final state in the LHCb kinematics obtained with the MRW and the KMR gluon uPDF are $\sigma_{\mathrm{MRW}} = 32.14 \; \mu b$ and $\sigma_{\mathrm{KMR}} = 58.85 \; \mu b$, respectively. It is therefore justified to conclude that the new $2 \to 2 + 3+ 4$ scheme with the KMR uPDF is more prefered by the low energy charm data, at least it leads to a better energy dependence of the charm cross section and it can be recommended as a better choice than usage of the MRW uPDF in any future studies.      
    
\subsection{Next-to-leading order collinear approach}

Having definite conclusions about the theoretical results obtained within the $k_{T}$-factorization, in the last step of the present analysis we move beyond this framework and apply the NLO collinear approximation. The corresponding results of the collinear approach may give another interesting point of reference and will make this study more complete. 

\begin{figure}[!h]
\begin{minipage}{0.47\textwidth}
  \centerline{\includegraphics[width=1.0\textwidth]{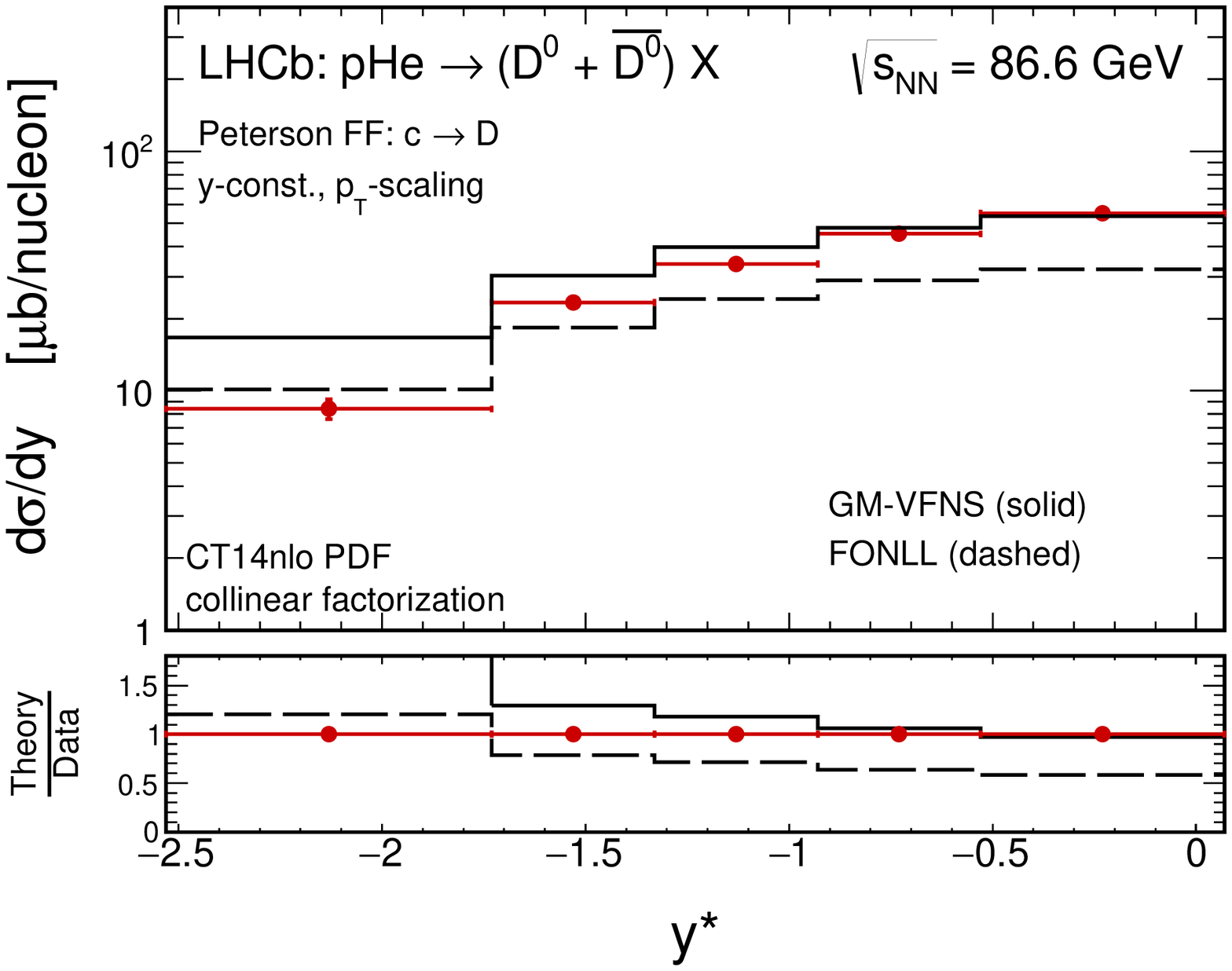}}
\end{minipage}
\begin{minipage}{0.47\textwidth}
  \centerline{\includegraphics[width=1.0\textwidth]{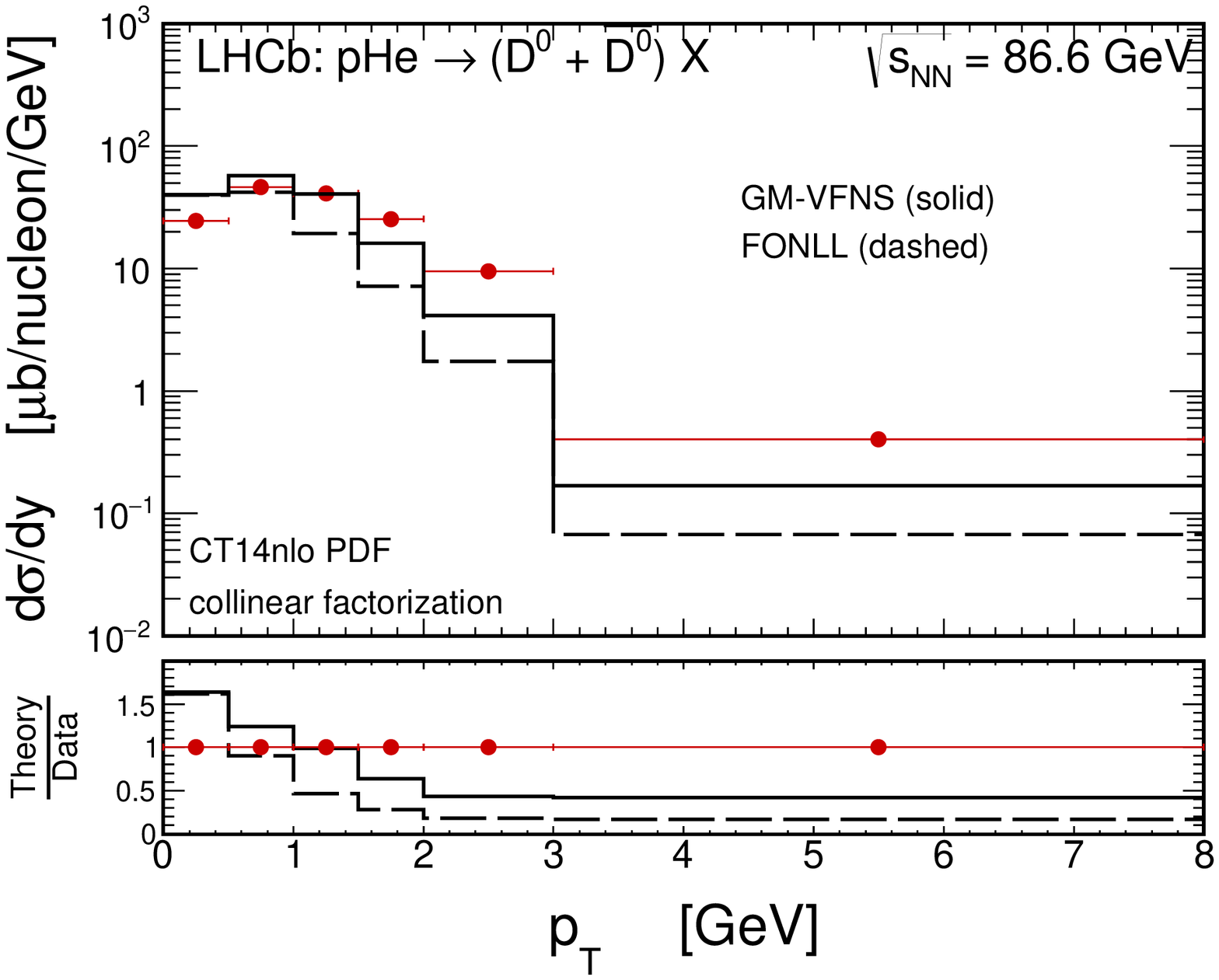}}
\end{minipage}
  \caption{
\small The transverse momentum distributions of $D^{0}$ meson (plus $\overline{D^{0}}$ antimeson)
for $p+^4\!\mathrm{He}$ collisions together with the LHCb data. Here the NLO collinear results for the FONLL (left) and GM-VFNS (right) frameworks are shown. Other details are specified in the figure.  
}
\label{fig:NLO}
\end{figure}

In Fig.~\ref{fig:NLO} we show rapitiy (left) and transverse momentum (right) distributions of $D^{0}$ meson measured in the LHCb fixed-target experiment together with the predictions of the FONLL \cite{Cacciari:1998it,Cacciari:2001td} (dashed lines) and the GM-VFNS \cite{Kniehl:2015fla,Benzke:2017yjn} (solid lines) frameworks.
The FONLL central prediction significantly underestimates the data points in the region of meson transverse momenta $ p_{T} > 1$ GeV. Visibly, some missing strength
is found with respect to both the experimental data set and to the $k_{T}$-factorization results presented in previous subsections. The GM-VFNS prediction is slightly closer to the data, however, it is received for $m_{c}= 1.3$ GeV while the FONLL result is for $m_{c}= 1.5$ GeV.
Even within the smaller charm quark mass the missing strength at larger $p_{T}$'s is also obtained. Potentially, the NNLO collinear predictions could change the slope at larger meson $p_{T}$'s and improve the situation but the NNLO calculation of differential cross sections for charm quarks is not yet available. Up to now, the framework of the $k_{T}$-factorization as discussed in the present paper is the only available method to study charm differential cross sections beyond the next-to-leading order.
    
\section{Conclusions}

We have considered production of neutral open charm mesons at $\sqrt{s} = 86.6$ GeV for the LHCb fixed-target experiment.
The numerical predictions have been done in the $k_{T}$-factorization approach. Two different schemes of the calculations have been discussed
depending on the model of unintegrated parton densities in a proton used in the analysis. A very good agreement with the LHCb fixed-target open charm data
has been obtained for both of them. The JH-2013-set2 CCFM unintegrated parton density describes the data very well already at leading-order with
$g^*g^* \to c\bar c$ off-shell mechanism only. Predictions based on the PB-NLO-set1 uPDFs needs to be applied within the higher-order perturbative calculations in order to describe the experimental data at similar level of quality as in the CCFM case. Similar conclusions are drawn for the case of the MRW and the KMR uPDFs, respectively. We have explicitly shown that the new $2 \to 2 + 3+ 4$ scheme with the KMR uPDF is more preferred by the low energy charm data than the standard calculation with the MRW uPDF, frequently and successfully used at larger LHC energies.      

We have demonstrated differences between the usage of the CCFM- and the DGLAP-based unintegrated parton distribution functions in phenomenological studies performed in the $k_{T}$-factorization. Both approaches provide a reasonable
agreement between the theory and experimental results. Application of both, the CCFM and the DGLAP frameworks in the large-$x$ limit explored in the LHCb fixed-target experiment seems to be supported by the data on charm production.
We also have compared the $k_{T}$-factorization predictions to the NLO collinear calculations.

The theoretical analysis of the low energy LHCb open charm data presented here could be useful to constrain
further theoretical studies of $\nu_{\tau}$ neutrino and $\overline{\nu}_{\tau}$ antineutrino production in future fixed-target experiment SHiP (see \textit{e.g.} Ref.~\cite{Maciula:2019clb}) which is now in comprehensive design phase.  

\vskip+5mm
{\bf Acknowledgments}\\
The author is very indebted to Maria Vittoria Garzelli, Victor P. Goncalves and Antoni Szczurek for interesting discussions as well as to Michael Benzke
for generating and sharing GM-VFNS predictions included in the figures. This work has been supported by the Mainz Institute for Theoretical Physics (MITP) of the Cluster of Excellence PRISMA+ (Project ID 39083149).


\end{document}